\documentclass[aps,prb,floatfix,amsmath,amssymb,eqsecnum,nofootinbib,twocolumn,superscriptaddress]{revtex4-1}

\usepackage{graphicx}
\usepackage{dcolumn}
\usepackage{bm}
\usepackage{amsmath}    
\usepackage[abs]{overpic}
\usepackage{relsize}  

\def\ssm{{\mathsmaller{\mathsmaller{\mathsmaller{<}}}}}
\def\sbg{{\mathsmaller{\mathsmaller{\mathsmaller{>}}}}}
\def\thb{{\tilde{\hbar}}}
\def\dbar{{\bar{d}}}
\def\Dbar{{\bar{D}}}

\begin{document}

\title{From classical to quantum criticality }


\author{Daniel Podolsky}
\affiliation{Department of Physics, Technion, Haifa 32000, Israel}

\author{Efrat Shimshoni}
\affiliation{Department of Physics, Bar-Ilan University, Ramat-Gan 52900, Israel}

\author{Pietro Silvi}
\affiliation{Institut f\"{u}r Quanteninformationsverarbeitung, Universit\"{a}t Ulm, D-89069 Ulm, Germany}

\author{Simone Montangero}
\affiliation{Institut f\"{u}r Quanteninformationsverarbeitung, Universit\"{a}t Ulm, D-89069 Ulm, Germany}

\author{Tommaso Calarco}
\affiliation{Institut f\"{u}r Quanteninformationsverarbeitung, Universit\"{a}t Ulm, D-89069 Ulm, Germany}

\author{Giovanna Morigi}
\affiliation{Theoretische Physik, Universit\"{a}t des Saarlandes, D-66123 Saarbr\"{u}cken, Germany}

\author{Shmuel Fishman}
\affiliation{Department of Physics, Technion, Haifa 32000, Israel}

\date{\today}

\begin{abstract}
We study the crossover from classical to quantum phase transitions at zero temperature within the framework of $\phi^4$ theory.  The classical transition at zero temperature can be described by the Landau theory, turning into a quantum Ising transition with the addition of quantum fluctuations.  We perform a calculation of the transition line in the regime where the quantum fluctuations are weak.  The calculation is based on a renormalization group analysis of the crossover between classical and quantum transitions, and is well controlled even for space-time dimensionality $D$ below 4.  In particular, for $D=2$ we obtain an analytic expression for the transition line which is valid for a wide range of parameters, as confirmed by numerical calculations based on the Density Matrix Renormalization Group. This behavior could be tested by measuring the phase diagram of the linear-zigzag instability in systems of trapped ions or
repulsively-interacting dipoles.
\end{abstract}

\maketitle


\section{Introduction}
\label{sec:intro}

One of the fascinating aspects of critical phenomena in
statistical mechanics is the universality that characterizes the
behavior of thermodynamic functions close to a second order phase
transition \cite{Stanley,Ma,Cardy}. This behavior was first modeled phenomenologically by Landau, whose theory can be microscopically derived using a mean-field theoretical approach \cite{Landau}. The classical mean-field theory
predicts the critical exponents of thermodynamic quantities
for several systems undergoing a second-order phase transition. It
predicts, however, wrong exponents and scaling functions for
dimensions $D$ lower than $4$ due to the enhanced role of
fluctuations at lower dimensionality.  The renormalization group is required to find their correct values for \cite{Wilson} $D<4$.  At $T=0$,  in particular,
quantum fluctuations become relevant near quantum critical points
at low dimensions. Here, the critical behavior is usually
described in a field theoretical framework whose dimension $D$ is
related to the spatial dimension $d$ by $D=d+z$, where $z$ is the dynamical critical
exponent \cite{QPTbook}.   Existing studies
of quantum phase transitions in $D=1+1$ dimensions are usually
either based on mapping to well-known critical models that can be
performed by employing conformal field theory, and/or analyzed
with full numerical simulations \cite{Cardy,QPTbook,Henkel}.

In this paper we connect classical mean-field and quantum critical
behaviors in $D=d+1$ in a systematic fashion ({\em i.e.}~we restrict ourselves to models with $z=1$). 
To this end we
introduce a dimensionless parameter $\thb$, which quantifies the
strength of quantum fluctuations. Quite generally, the strength of quantum fluctuations $\thb$ can be
expressed in terms of a ratio between typical kinetic and
potential energy scales, respectively $U_K$ and $U_P$, such that \cite{SMF,DMRG:1}
\begin{eqnarray}
\tilde{\hbar}\sim\sqrt{\frac{U_K}{U_P}}\,. \label{tilde_hbar_def}
\end{eqnarray}
We then analyze how the location of the critical point in
parameter space varies when we smoothly tune $\thb$ from $\thb=0$,
where the classical mean-field description holds, to a small
finite value. This tuning may be performed by
changing physical parameters, {\em e.g.}~the linear density in a chain of
interacting
particles such as trapped ions~\cite{Birkl92}.

The theory is captured by the zero temperature partition function
$Z$, which can be expressed in the form of a path integral
\cite{QPTbook}:
\begin{eqnarray}\label{Z0}
Z=\int {\mathcal D}\phi~e^{-\tilde{S}[\phi]/\thb}\ ,
\end{eqnarray}
where $\phi$ is a real field and $\tilde{S}$ is the dimensionless
action, which is defined on the $D$-dimensional Euclidean space-time ($D=d+1$)
as
\begin{eqnarray}\label{S0}
\tilde{S}&=&\int d^D r
\left[\frac{1}{2}\left(\partial_\mu\phi\right)^2-\frac{\varepsilon}{2}\phi^2+ g\phi^4\right]\,.
\end{eqnarray}
Here $\mu\in\{1\ldots D \}$ and $g>0$.  Moreover, the components of the space-time vector ${\bf r}$ have been rescaled in order to make the action dimensionless and to fix the speed of sound to unity.
The action
is defined with an implicit cutoff at short distances
corresponding to an ultraviolet cutoff $\Lambda$ in momentum
space. For an array of interacting particles with inter-particle
distance $a$, for instance, $\Lambda=\pi/a$.

The action Eq.~(\ref{S0}) undergoes an Ising phase transition at a
critical value $\varepsilon_c$~\cite{QPTbook,SMF,SMFlong}.  For
$\varepsilon<\varepsilon_c$ the system is disordered,
$\langle\phi\rangle=0$, corresponding to a paramagnetic phase, whereas for
$\varepsilon>\varepsilon_c$  the system orders
and $\langle\phi\rangle\ne 0$, corresponding to a ferromagnetic phase. The value of $\varepsilon_c$ depends on the strength of the quantum fluctuations: For
$\thb=0$ the partition function Eq.~(\ref{Z0}) is solved exactly
by a saddle-point evaluation, yielding a classical mean-field
transition at $\varepsilon_c=0$. For $\thb>0$, on the other hand, the transition is
shifted to $\varepsilon_c>0$, and becomes of the quantum Ising universality class.
Our purpose is to study the crossover between classical and quantum phase transitions
for small values of $\thb$.

This paper is organized as follows. In Sec. \ref{sec:QCP} we
present the derivation of a scaling relation for $\varepsilon_c(\hbar)$
employing a renormalization group (RG) procedure. The result is
verified numerically for $D=2$ in Sec. \ref{sec:num} by means of a Density Matrix Renormalization Group (DMRG)
calculation. In Sec. \ref{Conclusions} we summarize our main
results and conclusions. In particular, we discuss possible
experimental systems, which can serve as testbeds for these predictions.

\section{Dependence of the critical point on quantum fluctuations}
\label{sec:QCP}

Our aim is to determine the boundary between the disordered and the ordered phase  in parameter space for the model described by Eq. \eqref{S0} and at $T=0$, as the effect of quantum fluctuations, $\thb$, is increased from zero.  In this section we calculate $\varepsilon_c(\thb)$ as a function of $\thb$. We first obtain a crude result using naive scaling and then obtain a more accurate result with the help of the Renormalization Group (RG), using Wilson's momentum shell integration \cite{Ma,Wilson}.  This is a standard problem of crossover \cite{FisherPfeuty,NelsonCrossover}, but our problem is simple enough that it can be analyzed in more detail.

For any space-time dimensionality $D=d+1$, the model in Eq. \eqref{S0} has a fixed point at $\thb= 0$ and $\varepsilon=0$.  This fixed point leads to scaling that is mean-field in nature.  To see this, consider a generalization of the model to include high order anharmonicities, such as $\phi^6$, $\phi^8$, etc, which were implicitly ignored in Eq. \eqref{S0} and which are present in realistic systems ({\em e.g.}~for chains of interacting particles \cite{FCCM}):
\begin{eqnarray}
\frac{1}{\thb} \tilde{S}&=&\frac{1}{\thb}\int d^D r
\left[\frac{1}{2}\left(\partial_\mu\phi\right)^2-\frac{\varepsilon}{2}\phi^2+ g\phi^4
\right.\nonumber\\ &\,& \qquad \left.
+g_6 \phi^6+g_8 \phi^8+\ldots\right]\,.
\end{eqnarray}
As $\thb\to 0$, the partition function can be evaluated using a saddle-point approximation. Then, the phase transition occurs at $\varepsilon=0$, and near the phase transition the value of the order parameter is small, of order $\sqrt{|\varepsilon|}$, such that high anharmonicities such as $\phi^6$ and higher can be neglected (provided that $g>0$).  This can also be seen by introducing a rescaled field $\varphi \equiv \phi/\sqrt{\thb}$,
\begin{eqnarray}
\frac{1}{\thb} \tilde{S}
&=&\int d^D r
\left[\frac{1}{2}\left(\partial_\mu\varphi\right)^2-\frac{\varepsilon}{2}\varphi^2+ g\thb\varphi^4
\right.\nonumber\\ &\,& \qquad \left.
+g_6\thb^2 \varphi^6+g_8\thb^3 \varphi^8+\ldots\right]\,.
\end{eqnarray}
In this form it is clear that all anharmonicities rescale to zero  in the limit $\thb\to 0$, with higher anharmonicities more strongly suppressed.  Therefore, for small but positive $\thb$ (and $g>0$), it is sufficient to consider quartic anharmonicities only, as done in Eq.~\eqref{S0} and through the rest of this paper.  This is in contrast with scaling near the tricritical point, where the quartic term strictly vanishes and one must keep the sixth order term
\cite{Cardy}.

For positive $\thb$, fluctuations tend to extend the size of the disordered phase, thus shifting the phase transition to positive values of $\varepsilon$.  Provided $2\le D <4$ the resulting fixed point is in the Ising universality class.  However, for small $\thb$ the shape of the phase transition line, $\varepsilon_c(\thb)$, is expected to be controlled by scaling near the mean-field fixed point at $\thb=0=\varepsilon$.  We will use this fact to first give a naive derivation of $\varepsilon_c(\thb)$ and then improve on this estimate using the RG.

\subsection{Naive scaling and fluctuation corrections}

We proceed with a simple analysis based on spatial rescaling. For
this purpose we introduce the scaling factor $b>1$ and perform the
following rescaling of the space-time coordinate ${\bf r}$ as
\begin{eqnarray}
{\bf r}'={\bf r}/b\,,
\label{rescale}
\end{eqnarray}
which leads to the rescaling of the correlation length
\begin{eqnarray}
\xi'=\xi/b\,.
\end{eqnarray}
Using the critical scaling of the correlation length $\xi$ and of the order parameter $\phi$ near the mean-field fixed point at $\thb=0$ and $\varepsilon=0$,
\begin{eqnarray}
&&\xi\sim \varepsilon^{-\nu}\,,
\label{nu_def}\\
&&\phi \sim \varepsilon^\beta\;,
\label{beta_def}
\end{eqnarray}
one obtains
\begin{eqnarray}
\varepsilon'=\varepsilon \,b^{1/\nu}, \qquad    \phi'=\phi \,b^{\beta/\nu}\,,
\label{rescale2}
\end{eqnarray}
where the second equality directly follows from the first.  The critical exponents appearing in these expressions are the mean field exponents $\beta=\nu=\frac{1}{2}$. Then, all terms in the action [Eq. (\ref{S0})] are rescaled by the factor $b^{D-4}$,
\begin{eqnarray}
\tilde{S}=b^{D-4}\tilde{S}'\; ,
\end{eqnarray}
where $\tilde{S}'$ is obtained from $\tilde{S}$ by replacing the various variables with the corresponding primed ones. In the rescaling process, the expression for the partition function is unaffected, only the overall value of the action is changed. This can be summarized by rescaling the effective $\thb$,
\begin{eqnarray}
\thb'=\thb b^{4-D}\; ,
\end{eqnarray}
and shows that, by continuing the process, one moves farther away
from the mean-field critical point. (As a side remark, note that
$D=4$ is a marginal case of the rescaling and that for $D>4$
quantum fluctuations become irrelevant.)

Let us consider a reference point $(\varepsilon^*,\thb^*)$ lying on the phase transition line.  We can then locate other points on the transition line that lie closer to the mean-field critical point and which reach the reference point after an iterative repetition of this rescaling.  Let $(\varepsilon,\thb)$ be one such point.  Assuming that it takes $l$ rescaling steps to reach the reference point, this yields $\thb^*=\thb b^{(4-D)l}$ and $\varepsilon^*=\varepsilon b^{2l}$, where in the last equation we used Eq. (\ref{nu_def}) with $\nu=1/2$.  This implies that
\begin{eqnarray}
\thb\sim \varepsilon^{(4-D)/2}\,.
\label{naive}
\end{eqnarray}

\begin{figure}[ht]
\begin{center}
\includegraphics[width=0.375\textwidth]{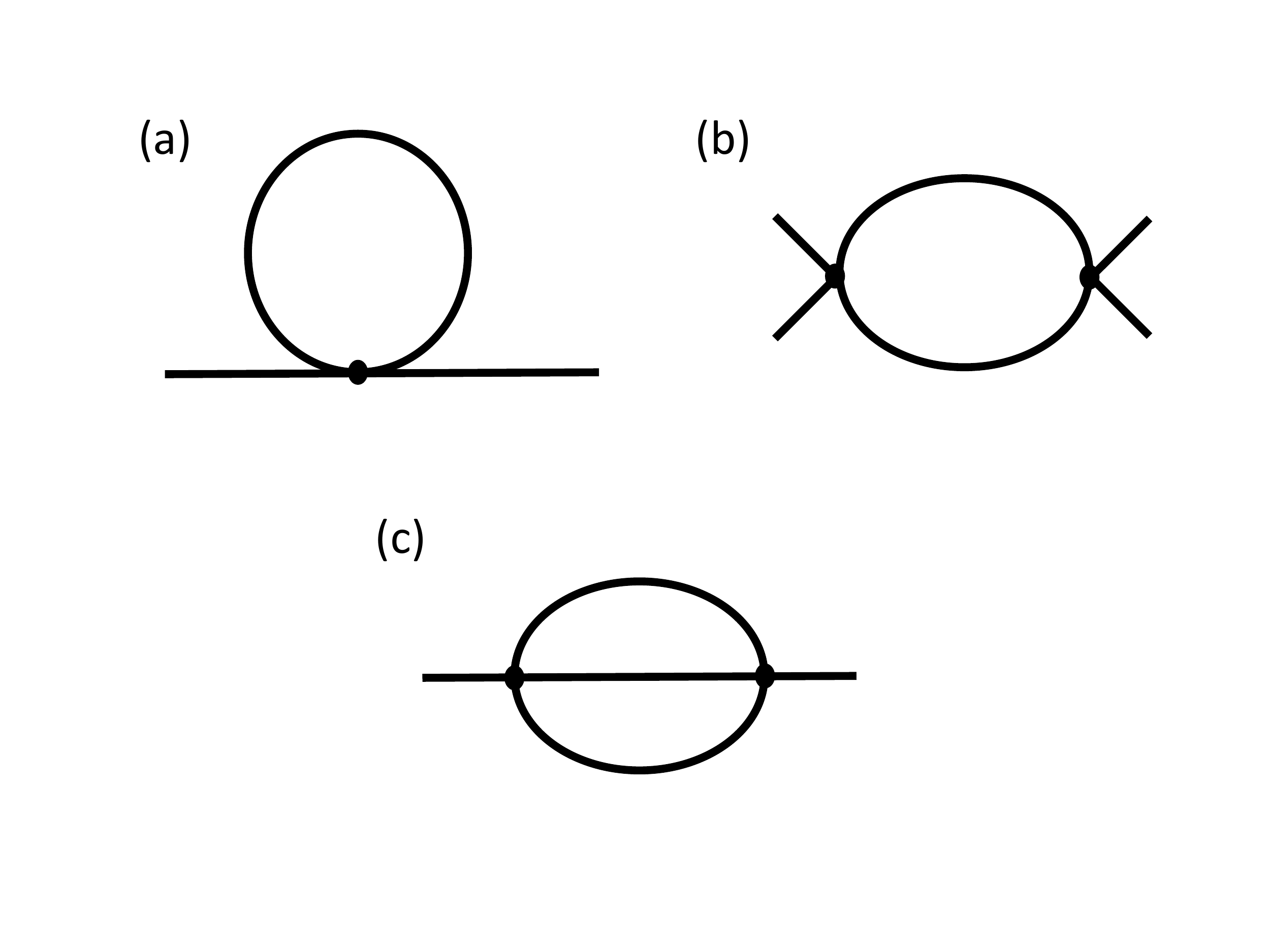}
\caption{(a) and (c) Lowest order Feynman diagrams contributing to the renormalization of $\varepsilon$ and (b) to the renormalization of $g$.}\label{fig:diagram}
\end{center}
\end{figure}

Let us now explore the effects of fluctuations.  For small values of $\thb$, the leading correction to equation Eq.~(\ref{naive}) comes from the diagram in Fig.~\ref{fig:diagram}.(a), which leads to a self-consistent equation for the renormalization of $\varepsilon$:
\begin{eqnarray}
\varepsilon_{\rm ren}&=&\varepsilon-12 g\thb\!\!\!\!\int\limits_{|{\bf q}|<\Lambda}\!\!\!\!\frac{{\rm d}^D  q}{(2\pi)^D}\frac{1}{q^2-\varepsilon_{\rm ren}}\label{eq:eps1}\,,
\end{eqnarray}
where $q$ is the modulus of the Euclidean momentum $D$-vector
${\bf q}$. Although the physical cutoff only applies to the
space-like components of the momentum, we have simplified the integral by taking a
space-time symmetric cutoff $|{\bf q}| < \Lambda$.  This is
expected not to affect any universal properties arising from the
infrared divergences \cite{Ma,Wilson}. The critical value $\varepsilon_c$ is then
found by setting $\varepsilon_{\rm ren}=0$.  For $D>2$, the
integral above is convergent at $\varepsilon_{\rm ren}$=0, giving
\begin{eqnarray}
\varepsilon_c&=&12 g\thb\!\!\!\!\int\limits_{|{\bf q}|<\Lambda}\!\!\!\!\frac{{\rm d}^D  q}{(2\pi)^D}\frac{1}{q^2}=\frac{A_D}{D-2} g\thb\label{eq:eps2}
\end{eqnarray}
where
\begin{eqnarray}
A_D=\frac{12 S_{D-1}\Lambda^{D-2}}{(2\pi)^D}
\end{eqnarray}
and $S_d$ is the area of a $d$-sphere of unit radius ($S_1=2\pi$, $S_2=4\pi$, and $S_3=2\pi^2$).  Hence, for $D>2$, $\varepsilon_c$ is linear in $\thb$.  This result seems to contradict Eq.~(\ref{naive}).  This can be understood from the fact that the critical value $\varepsilon_c$ itself is not
a universal number -- the correction in Eq.~(\ref{eq:eps2}) constitutes an {\em analytic} shift in the critical point by a non-universal amount. However, as we will show in the following, if we introduce a shifted variable $R$, defined by
\begin{eqnarray}
R\equiv\varepsilon-\frac{A_D}{D-2} g \label{eq:R}\thb
\end{eqnarray}
then  $R$ and $\thb$ satisfy the naive scaling for $2<D<4$, namely,
\begin{eqnarray}
R_c \sim \thb^{2/(4-D)}\,.
\end{eqnarray}
On the other hand, for $D=2$ the integral in Eq.~(\ref{eq:eps1}) is logarithmically divergent as $\varepsilon_{\rm ren}\to 0$.  In this case the infrared fluctuations renormalize $\varepsilon$ in a more fundamental way, requiring a more careful treatment.

\subsection{Renormalization group}

In the following fluctuations will be taken into account using the Renormalization Group (RG), by implementing a momentum-shell integration \cite{Ma,Wilson}. For this purpose, we introduce the Fourier transform  $\phi({\bf q})$ of the real field $\phi({\bf r})$, such that
\begin{eqnarray}
\phi({\bf r})=\int\limits_{|{\bf q}|<\Lambda}\frac{{\rm d}^D
q}{(2\pi)^D} \phi({\bf q})\exp(-{\rm i}({\bf q}\cdot {\bf r}))\,.
\end{eqnarray}
Then Eq.~(\ref{S0}) can be recast in the form
\begin{eqnarray} \label{action_q}
\tilde{S}&=&\frac{1}{2} \int\limits_q  \left({\bf q}^2-\varepsilon\right)\phi({\bf q})\phi(-{\bf q})+\\
&&\, g\int\limits_{q_1}\int\limits_{q_2}\int\limits_{q_3}
\phi({\bf q}_1)
\phi({\bf q}_2)
\phi({\bf q}_3)
\phi(-{\bf q}_1-{\bf q}_2-{\bf q}_3)\,,\nonumber
\end{eqnarray}
where $\int_q := \int_{|{\bf q}|<\Lambda} {\rm d}^Dq/(2\pi)^D$ .

We now define a momentum shell $[ \Lambda/b,\Lambda]$, with $b>1$. This momentum shell identifies a partition, so that we write $\phi({\bf q})=\phi_\ssm({\bf q})+\phi_\sbg({\bf q})$,  where $\phi_\ssm$ and $\phi_\sbg$ only have support outside the shell (for small momenta) and within the momentum shell (for large momenta), respectively.
In detail,
\begin{eqnarray}
\phi_\ssm({\bf q})=\left\{\begin{array}{l l}\phi({\bf q}) & \mathrm{for}\,\ |{\bf q}|<\Lambda/b \\
0 & \mathrm{otherwise}\end{array}\right.
\end{eqnarray}
and
\begin{eqnarray}
\phi_\sbg({\bf q})=\left\{\begin{array}{l l}\phi({\bf q}) & \mathrm{for}\,\ \Lambda/b<|{\bf q}|<\Lambda  \\
0 & \mathrm{otherwise}\end{array}\right.
\end{eqnarray}
In terms of these fields, the partition function can be written schematically as
\begin{eqnarray}
Z&=&\int {\mathcal{D}}\phi_\ssm\,{\mathcal{D}}\phi_\sbg \,e^{-\tilde{S}\left[\phi_\ssm,\phi_\sbg \right]/\thb}\\
\tilde{S} &=&\frac{1}{2} \int\limits_q (q^2-\varepsilon)\left[ \phi_\ssm^2+\phi_\sbg^2\right]\\
&\,&+g\int\limits_{q_1,q_2,q_3}\left[\phi_\ssm^4+4\phi_\ssm^3\phi_\sbg+6\phi_\ssm^2\phi_\sbg^2+4\phi_\ssm\phi_\sbg^3+\phi_\sbg^4\right].\nonumber
\end{eqnarray}
where the cross term $\phi_\ssm \phi_\sbg$ vanishes due to momentum conservation since the two fields do not have common support in the Fourier domain.

The RG procedure consists of two steps.  In the first step, the high-frequency field $\phi_\sbg$ is integrated out, yielding an effective action $\tilde{S}_{\rm eff}[\phi_\ssm]$ involving only the small momentum field, which is defined by the equation
\begin{eqnarray}
e^{-\tilde{S}_{\rm eff}[\phi_\ssm]/\thb}=\int {\mathcal{D}}\phi_\sbg e^{-\tilde{S}[\phi_\ssm,\phi_\sbg]/\thb}\,.
\end{eqnarray}
The resulting field theory has a reduced cutoff $\Lambda/b$.  In the second step, all length scales are rescaled by a factor of $b$ following Eq.~(\ref{rescale}), such that the cutoff is returned to its original value $\Lambda$.

For small $\thb$ the RG is controlled by Feynman diagrams with few loops. We will work to one loop order, taking into account the diagrams in Fig.~\ref{fig:diagram}.(a) and (b).  These diagrams lead to renormalization of $\varepsilon$ and $g$, respectively.  The diagram in Fig.~\ref{fig:diagram}.(c), contributing to the field renormalization, is of higher order in $\thb$ and will not be taken into account in this analysis.

Let us first consider diagram \ref{fig:diagram}.(a). It describes fluctuations that originate from the $\phi_\ssm^2\phi_\sbg^2$ interaction.  Upon integrating out $\phi_\sbg$, this leads to a renormalization of $\varepsilon$,
\begin{eqnarray}
\varepsilon_{\rm eff}&=&\varepsilon-12 g\thb\!\!\!\!\int\limits_{\Lambda/b<|{\bf q}|<\Lambda}\!\!\!\!\frac{{\rm d}^D  q}{(2\pi)^D}\frac{1}{q^2-\varepsilon}\label{eq:eps}\\
&=&\varepsilon-\thb g \frac{A_D}{1-\varepsilon/\Lambda^2}d\ell\nonumber
\end{eqnarray}
where we have taken the limit of a thin momentum shell, $b=1+d\ell$.  Similarly, diagram \ref{fig:diagram}(b) (together with two additional diagrams related to it by crossing symmetry) leads to a renormalization of $g$,
\begin{eqnarray}
g_{\rm eff}&=&g-36 g^2\thb\!\!\!\!\int\limits_{\Lambda/b<|{\bf q}|<\Lambda}\!\!\!\!\frac{{\rm d}^D  q}{(2\pi)^D}\frac{1}{(q^2-\varepsilon)^2}\label{eq:eps3}\\
&=&g-\thb g^2\frac{B_D}{(1-\varepsilon/\Lambda^2)^2}d\ell\nonumber
\end{eqnarray}
where $B_D=3A_D/\Lambda^2$.

We now perform a rescaling of distances and fields, according to Eq.~(\ref{rescale}).  This yields an action with the same form as the original, including the same cutoff, but with the rescaled parameters,
\begin{eqnarray}
\varepsilon' &=& b^2\varepsilon_{\rm eff}\,,\\
\thb'&=&b^{4-D} \thb\,,\\
g'&=&g_{\rm eff} \,.
\end{eqnarray}
The corresponding differential flow equations read
\begin{eqnarray}
\frac{d \varepsilon}{d\ell} &=& 2\varepsilon-\thb g \frac{A_D}{1-\varepsilon/\Lambda^2}\,,\\
\frac{d\thb}{ d\ell}&=& (4-D) \thb\,,\\
\frac{dg}{d\ell}&=&-\thb g^2\frac{B_D}{(1-\varepsilon/\Lambda^2)^2}\,.
\end{eqnarray}
Note that at this order in the RG, the flow equation for $\thb$ is trivial (this would no longer be the case after the inclusion of diagram \ref{fig:diagram}.(c)).  Thus, it is useful to reduce the number of flow equations by introducing the parameter $u\equiv \thb g$ ,
\begin{eqnarray}
\frac{d \varepsilon}{d\ell} &=& 2\varepsilon-u \frac{A_D}{1-\varepsilon/\Lambda^2}\,,\\
\frac{du}{d\ell}&=&(4-D)u-u^2\frac{B_D}{(1-\varepsilon/\Lambda^2)^2}\,.  \label{eq:RG1loop}
\end{eqnarray}
These equations are exact for small values of $\varepsilon$ and $u$.  Hence, the initial RG flow near the mean-field fixed point  is captured by these equations, regardless of dimensionality.  For $2\le D<4$, these equations predict an additional, Ising, fixed point at $\varepsilon_I=\Lambda^2\frac{4-D}{10-D}$ and $u_I=\frac{4-D}{B_D}\left(\frac{6}{10-D}\right)^2$.  The location of this fixed point is not reliable, except for the special case in which $D$ is close to four dimensions.  Then, $\varepsilon_I$ and $u_I$ are small and the entire RG flow is well controlled by these equations.

The phase transition line is given by the set of points that flows to the Ising critical point.  Suppose $(\varepsilon^*,u^*)$ is one such point, but which lies close enough to the mean-field point that non-linear terms such as $(u^*)^2$ are small.  Then, for regions of the phase transition closer to the mean-field fixed point than $(\varepsilon^*,u^*)$, we can find the phase transition line by solving the linearized equations,
\begin{eqnarray} 
\frac{d \varepsilon}{d\ell} &=& 2\varepsilon-u A_D\,,\label{eq:RGlinear1}\\
\frac{du}{d\ell}&=&(4-D)u
\,, \label{eq:RGlinear2}
\end{eqnarray}
and searching for the set of points that reach the reference point $(\varepsilon^*,u^*)$.  These equations are readily solved to yield $\ell$-dependent coupling constants.  In what follows, we will consider the cases $D=2$ and $2<D<4$ separately.

\subsubsection{Two dimensions}

For $D=2$, the solution to the linearized equations is
\begin{eqnarray}
\varepsilon_\ell &=& \left(\varepsilon-\frac{6 u}{ \pi}\ell \right)e^{2\ell}\,,\\
u_\ell&=& u\, e^{2\ell}\,,
\end{eqnarray}
where we used $A_2=6/\pi$, and where the variables $\varepsilon$ and $u$ without subscript denote the physical (bare) values.  Let us suppose that $\varepsilon$ and $u$ are points on the transition line, but which lie much closer to the mean-field point than the reference point.  Then the flow runs for a long RG time $\ell^*$ before the renormalized parameters reach the reference point, $\varepsilon_{\ell^*}=\varepsilon^*$ and $u_{\ell^*}=u^*$.  This gives
\begin{eqnarray}
\ell^* &=& \frac{1}{2}\ln \frac{u^*}{ u}\,,\\
\varepsilon^* e^{-2\ell^*} &=& \varepsilon-\frac{3 u}{ \pi}\ln \frac{u^*}{ u}\,.
\label{Eq:e0}
\end{eqnarray}
In particular, as $(\varepsilon,u)$ approach the mean-field point, corresponding to the limit $\ell^*\to\infty$, the left-hand side of Eq.   \eqref{Eq:e0} tends to zero. Thus, near the mean-field point the bare values of $\varepsilon$ and $u$ satisfy the relation
\begin{eqnarray}
\varepsilon_c=\frac{3 u}{ \pi}\,\left|\ln \frac{u}{ u^*}\right|\label{eq:ulog}
\end{eqnarray}
that is asymptotically exact and universal as $u\to 0$.
This formula shows that there are logarithmic corrections to the linear dependence, which we obtained through naive scaling in Eq. \eqref{naive}. Note that the value of $u^*$ in this result is ambiguous and non-universal.  Using the relation $u=\thb g$ in Eq.~\eqref{eq:ulog}, delivers the relation between $\varepsilon_c$ and $\thb$,
\begin{eqnarray} \label{eq:hlogh}
\varepsilon_c\sim\frac{3 g\thb}{ \pi}\,\left(\left|\ln \thb\right|-\left|\ln \thb^*\right|\right)\ \label{eq:varCritical2d}
\end{eqnarray}
which is asymptotically exact in the limit of small $\thb$.  Here, $\thb^*$ is a system-dependent parameter.  Note that the leading term in Eq.~\eqref{eq:varCritical2d} is universal, and that it is dominant as $\thb\to 0$.

\subsubsection{Dimension $2<D<4$}

\begin{figure}[ht]
\begin{center}
\begin{overpic}[width = 0.375\textwidth]{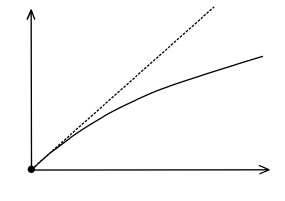}
  \put(0, 100){\large $\varepsilon_c$}
  \put(160, 0){\large $\thb$}
 \end{overpic}
 \end{center}
\caption{Phase diagram for $D=3$ in the case $k<0$, see Eq.~(\ref{eq:Rscaling}).  The solid line denotes the transition line, whereas the dashed line corresponds to $R=0$.  The classical critical point is located at the origin.\label{fig:pd3d}}
\end{figure}

For $2<D<4$ dimensions, the solutions of the linearized equations (\ref{eq:RGlinear1}) and \eqref{eq:RGlinear2} are
\begin{eqnarray}
\varepsilon_\ell-\frac{A_D}{D-2} u_\ell&=&\left(\varepsilon-\frac{A_D}{D-2} u\right)e^{2\ell}\,,\\
u_\ell &=& u e^{(4-D)\ell}\,.
\end{eqnarray}
As before, we consider bare parameters $\varepsilon$ and $u$ lying on the transition line.  If these are small, it takes a long RG time $\ell^*$ to reach the reference point $(\varepsilon^*,u^*)$.  This yields,
\begin{eqnarray}
R^*&=&R\, e^{2\ell^*}\,,\\
u^* &=& u\, e^{(4-D)\ell^*}\,,
\end{eqnarray}
where the shifted parameter $R$ was defined in Eq.~(\ref{eq:R}).  Solving for $\ell^*$ yields a direct relation between $R$ and $u$ at the transition line,
\begin{eqnarray}
R_c=k\, u^{2/(4-D)}
\end{eqnarray}
where $k=R^*/(u^*)^{2/(4-D)}$ is a non-universal constant.  Hence, we see that the shifted parameter $R$ satisfies the naive scaling, as expected.  In terms of $\varepsilon$ and $\thb$, this result reads,
\begin{eqnarray}
\varepsilon_c=\frac{A_D}{D-2} g\thb + k (g\thb)^{2/(4-D)} \,.
\label{eq:Rscaling}
\end{eqnarray}
This is shown schematically in Fig.~\ref{fig:pd3d}.

Finally, in order to fix the sign of the coefficient $k$, we must determine whether the Ising fixed point lies above or below the line $R=0$. This can be determined by computing $\varepsilon_c$ to order $u^2$, thus including the diagram in Fig.~\ref{fig:diagram}.(c).  This gives a negative contribution to $\varepsilon_c$, implying that $k<0$.\cite{TenenbaumKatanHiggs}

In summary, we analyzed the crossover from mean-field behavior to quantum critical behavior at the fixed point $\varepsilon=0$ and $\thb=0$.  Both $\varepsilon$ and $\thb$ are relevant variables with stability exponents
\begin{eqnarray}
\lambda_\varepsilon&=&2\,,\\
\lambda_\thb&=&4-D\,,
\end{eqnarray}
resulting in the crossover exponent $\Phi=(4-D)/2$.\cite{FisherPfeuty,NelsonCrossover}  Within the standard theory the transition line is
\begin{eqnarray}
R_c\sim \thb^{1/\Phi}\,.
\end{eqnarray}
For $D=2$, this simplistic scaling fails and Eq.~(\ref{eq:varCritical2d}) holds instead, which involves a logarithmic correction to the standard result.

\section{Numerical Analysis}
\label{sec:num}

In order to check these predictions numerically, we resort to simulations based on the
DMRG~\cite{White92}, a numerical technique designed for studying
quantum many-body systems in one-dimensional lattices~\cite{PrimoMPS}. We employ a DMRG algorithm developed to simulate a lattice version of Eq.~(\ref{S0}) for $D=1+1$, defined
by the Hamiltonian
\begin{eqnarray} \label{eq:hamDisc}
\mathcal H = \frac{1}{2} \sum_{j = 1}^{L} \left[\pi_j^2 - \varepsilon \phi_j^2 + \left( \phi_j -\phi_{j+1} \right)^2 + 2g\phi_j^4 \right]\,,
\end{eqnarray}
with $L$ the number of lattice sites and $\phi_j$, $\pi_j$ conjugate variables satisfying the commutation relation
\begin{eqnarray} \label{eq:gidef}
 [\phi_j,\pi_\ell] ={\rm  i} \thb \delta_{j,\ell}\,.
\end{eqnarray}
By means of this algorithm it was verified that the transition described by the Hamiltonian \eqref{eq:hamDisc}  is in the $D=1+1$ Ising universality class~\cite{DMRG:1,DMRG:2}.

In the algorithm, we adopt a local space truncation approach that first appeared in Ref.~\onlinecite{iblisdir07}.
In this approach, we define a related single-particle quantum problem for the local Hamiltonian and find its $\dbar$ lowest-energy eigenfunctions $|\psi_q\rangle$, which then form a local basis $\{|\psi_q\rangle \}_{q = 1, \ldots,\dbar}$ for the many-body problem. In order to keep track of the error generated by truncating the basis, we perform several trials of the same problem for various values of $\dbar = 2 \ldots 100$, and check convergence of the outcomes. Furthermore, we keep track of the populations of the local basis levels on each site and verify that the occupation probabilities in the one-site reduced density matrix
decrease exponentially with the level index $q$.  Via a standard DMRG procedure \cite{White92} we search for the ground state of the Hamiltonian \eqref{eq:hamDisc}, expressed in the local basis, for finite system size $L$ with Open Boundary Conditions (OBC). The latter choice is due to a natural aptitude of DMRG towards OBC: in this scenario it converges faster, and has enhanced precision and stability. In the various simulation runs that we considered, we found that a local basis dimension $\dbar = 30$ and a DMRG bondlink $\Dbar = 50$ were sufficient to guarantee convergence of the results within our precision threshold (typically $10^{-10}$).

In order to determine the transition line, we compute the ground state for a set of points ($\thb,\varepsilon$).  This is performed for a variety of system sizes $L$, typically
pushing up to $L=3000$ sites. For each simulation, we obtain the order
parameter $\langle\phi\rangle_L(\thb, \varepsilon)$ using a procedure outlined in Ref.~\onlinecite{DMRG:1}. Finally, we extrapolate to the thermodynamic limit
$\langle\phi\rangle_{\infty} = \lim_{L \to \infty} \langle\phi\rangle_L$ using finite size scaling and
discriminate whether its value is zero or not, allowing us to determine the
phase boundary. The order parameter is typically a very smooth function
of the size, and we find the procedure of locating the transition to be robust.  The error bars were fixed conservatively, and are dominated by the limitations in determining the vanishing point of the order parameter.

\begin{figure}
 \begin{center}
 \begin{overpic}[width = \columnwidth, unit=1pt]{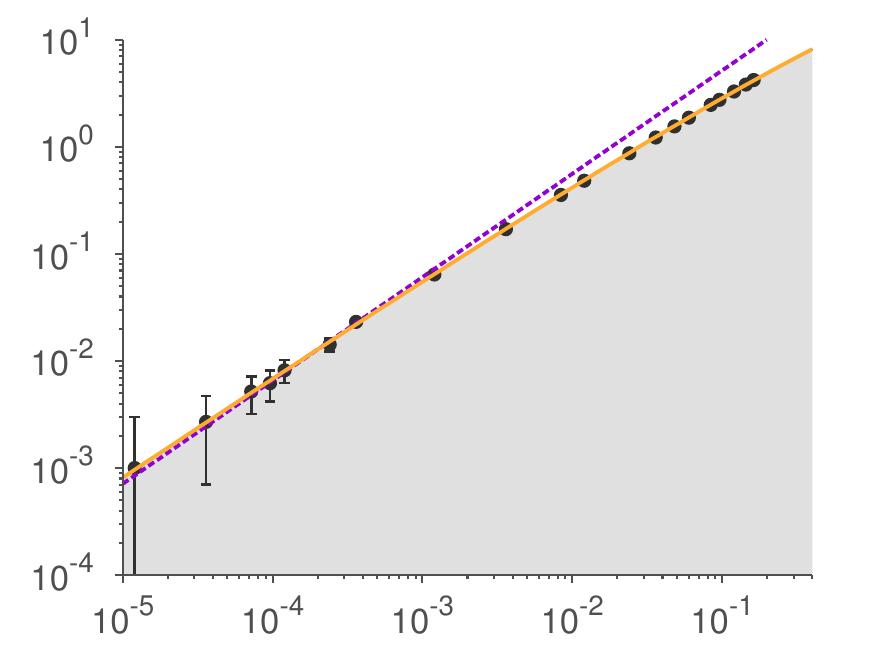}
  \put(0, 95){\large $\varepsilon$}
  \put(178, 0){\large $\thb$}
  \put(55, 150){\large $\langle\phi\rangle \neq 0$}
  \put(166, 50){\large $\langle\phi\rangle = 0$}
 \end{overpic}
 \end{center}
\caption{ \label{fig:DMRG}
(color online) Phase diagram in the
$(\thb, \varepsilon)$ parameter space for $g \simeq 6.027$.
The black dots are the critical points $\varepsilon_c(\thb)$ calculated via DMRG.
The solid orange curve is a fit to Eq.~\eqref{eq:hlogh} with $|\ln \thb^*| = 2.632 \pm 0.008$. 
The dashed violet curve is a power-law fit to Eq.~\eqref{eq:powerlaw} restricted to data in the
interval $[10^{-5},3\times 10^{-4}]$, with fitted exponent $\zeta=0.97 \pm 0.05$.
}
\end{figure}
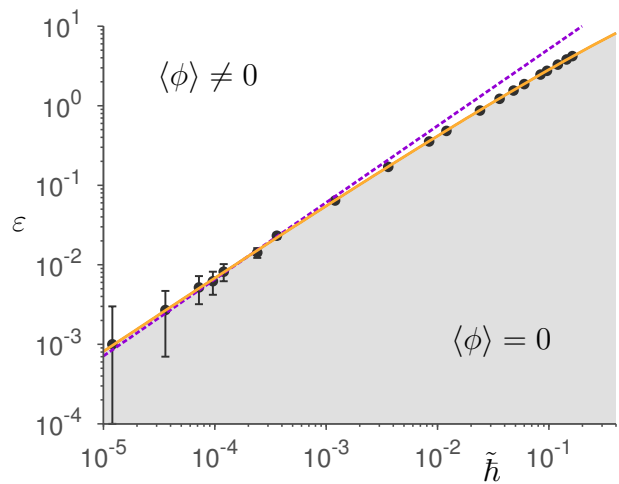

The value $\varepsilon_c(\thb)$ at which the phase transition
occurs has been located for several $\thb$
in the range $10^{-5} \leq \thb \leq 0.2$.  The results are shown in Fig.~\ref{fig:DMRG}.
A critical curve separates the ordered phase (white area),
from the disordered phase (grey area).  
As expected, as
$\thb$ increases, the magnitude of quantum fluctuations increases
as well, and as a result the order is destroyed. Hence, $\varepsilon_c(\thb)$ is a
monotonically increasing function of $\thb$.
Within a very narrow range $\thb\in [10^{-5},3\times 10^{-4}]$, the data can be fit to a power law
\begin{eqnarray}
\varepsilon_c\sim |\thb|^\zeta\,,
\label{eq:powerlaw}
\end{eqnarray}
as shown by the dashed violet line.  
The fitted exponent $\zeta = 0.97 \pm 0.05$ is consistent with the naive scaling in Eq.~\eqref{naive}. However, an excellent fit to
a much wider range  $\thb\in [10^{-5},0.2]$ is obtained by fitting to Eq.~\eqref{eq:hlogh} (solid orange curve).  Note that the fit involves only
one free parameter, $\thb^*$, which is found to be $\thb^*\approx 0.07$.  As a consistency check of the formalism, one can verify that the value of $g\hbar^*\approx 0.4$ is sufficiently small for the linearized RG equations, Eqs.~\eqref{eq:RGlinear1} and \eqref{eq:RGlinear2}, to be justified, assuming $\Lambda\approx \pi$.

\section{Discussion}
\label{Conclusions}  We have shown that the crossover from
classical to quantum phase transitions, described by a $\phi^4$
model, can be studied using Wilson's renormalization group around
the fixed point of mean-field theory. This results in the
prediction of the transition line at weak quantum fluctuations, $\hbar \to
0$, which contains a universal feature that can be calculated even for dimensions 
well below 4. In particular, for $D=2$ (corresponding to one space dimension) we
derive an expression relating the transition point $\varepsilon_c$
to the effective strength of the quantum fluctuations $\thb$ [Eq.
(\ref{eq:varCritical2d})] which is asymptotically exact in the
limit of small $\thb$. This analytical prediction is confirmed over four orders of magnitude by
a DMRG numerical calculation.

A physical problem described by this theory is the structural
transition of repulsively interacting particles from a linear
chain to a zigzag structure. Examples include electrons in
nanowires~\cite{MML,JM2013}, ultracold dipolar gases in elongated
traps~\cite{Astrakharchik,Lahaye,Altman,DMRG:2}, vortex lines in Bose-Einstein
condensates~\cite{Santos,Busch,DMRG:2}, and ion Coulomb crystals in
traps~\cite{Birkl92,Dubin93,Schiffer93,FCCM,SMFlong,DMRG:2}.  These different physical systems correspond to different strengths of the potential energy
$U_P$ relative to quantum fluctuations provided by the kinetic
energy $U_K$, and thus to different values of $\thb$ [Eq.
(\ref{tilde_hbar_def})]. In particular, due to the large particle
mass and the strong repulsive Coulomb interaction, ion Coulomb
crystals are characterized by values $\thb\ll 1$, namely, the
effect of quantum fluctuations is typically very small. The
transition they undergo from a linear to a zigzag structure can thus be
a setting for experimentally characterizing the crossover
from classical to quantum phase transitions. Here, $\thb$ is tuned
by the linear inter-particle distance $a$, the ion mass $m$, and
its charge $Q$ by determining the kinetic energy
$U_K=\hbar^2/(2ma^2)$ and the potential energy $U_P=Q^2/a$. The
transition from a linear to a zigzag structure can be tuned by
continuously varying either the ion density ($1/a$) or the
frequency of the transverse harmonic trap $\nu_t$, which dictate
the parameter $\varepsilon$ in Eq. (\ref{S0}) via
\cite{SMF,SMFlong}
\begin{eqnarray}
\varepsilon=\frac{ma^2(\nu_c^2-\nu_t^2)}{U_P} \label{eps_def}
\end{eqnarray}
with $\nu_c^2=\frac{7}{2}\frac{Q^2}{ma^3}\zeta(3)$. In this system, the
determination of the transition point $\varepsilon_c$ could be
performed by means of Bragg spectroscopy; estimates of the
experimental parameter regimes can be found in Ref.
\onlinecite{SMFlong}.

\acknowledgements
The authors gratefully acknowledge John Cardy and Malte Henkel for stimulating and illuminating discussions and helpful comments.  This work was supported by grants from the Israel Science Foundation (ISF), US-Israel Binational Science Foundation (BSF), the European Union
under grant agreement no. 276923 -- MC-MOTIPROX, the EU project SIQS, the German Research Foundation (DFG, projects: SFB/TRR21, Heisenberg grant),  the Minerva Center for Complex Systems at the Technion, and the Shlomo Kaplansky Academic Chair.

\end{document}